\documentclass{PoS}

\usepackage[caption=false]{subfig}
\usepackage[numbers,sort&compress]{natbib}
\usepackage[T1]{fontenc}
\usepackage{lmodern}
\usepackage{soul}
\usepackage{enumitem}
\usepackage{xcolor}
\usepackage[titletoc]{appendix}
\usepackage{overpic}
\usepackage[normalem]{ulem}
\usepackage[nointegrals]{wasysym}

\usepackage[caption=false]{subfig}
\usepackage{graphicx}
\usepackage{nicefrac}
\usepackage{tabularx}
\usepackage{booktabs}
\usepackage{multirow}
\usepackage{amssymb,amsmath}
\usepackage{footnote}
\usepackage{bm}
\usepackage[capitalize]{cleveref}[2012/02/15]

\usepackage{accents}

\usepackage{tabu}

\usepackage{siunitx}
\DeclareSIUnit\parsec{pc}
\DeclareSIUnit\lightyear{ly}

\let\OldAng\ang%
\renewcommand*{\ang}[2][]{%
    \OldAng[scientific-notation=false,#1]{#2}%
}
\DeclareSIUnit\year{yr}
\DeclareSIUnit\erg{erg}
\DeclareSIUnit\msun{M_{\astrosun}}
\DeclareSIUnit{\GeV}{\giga\electronvolt}
\DeclareSIUnit{\TeV}{\tera\electronvolt}
\DeclareSIUnit{\PeV}{\peta\electronvolt}
\DeclareSIUnit{\MeV}{\mega\electronvolt}
\DeclareSIUnit{\eV}{\electronvolt}
\DeclareSIUnit{\smm}{\square\metre\second}
\DeclareSIUnit{\smmr}{\metre^{-2}\second^{-1}}
\DeclareSIUnit{\dc}{d.c.}
\DeclareSIUnit{\pe}{p.e.}
\DeclareSIUnit{\nucleon}{nucleon}

\newcommand{\de}{\mathrm{d}}

\definecolor{desyOrange}{RGB}{242,142,0}
\usepackage{csquotes}
\usepackage[subtle]{savetrees}
\title{Measurement of the Iron Spectrum in Cosmic Rays with VERITAS}

\ShortTitle{Iron Spectrum}

\author{\speaker{Henrike Fleischhack}{} for the VERITAS collaboration\\
        Michigan Technological University \\
        E-mail: \email{hfleisch@mtu.edu}}

\abstract{The elemental energy spectra of cosmic rays play an important role in understanding their acceleration and propagation. Most current results are obtained either from direct measurements by balloon or satellite detectors, or from indirect measurements by air shower detector arrays on the Earth's surface. Imaging Air Cherenkov Telescopes (IACTs), used primarily for gamma-ray astronomy, can also be used for cosmic-ray physics. They are able to measure Cherenkov light emitted both by heavy nuclei and by secondary particles produced in their air showers, and are thus sensitive to the charge and energy of cosmic ray particles with energies of tens to hundreds of TeV.

A measurement of the energy spectrum of iron nuclei, based on 71 hours of data taken by the VERITAS array of IACTs between 2009 and 2012, will be presented. The energy and other properties of the primary particle are reconstructed using a template-based likelihood fit. The event selection makes use of direct Cherenkov light, which is emitted by the primary particle before starting the air shower. A multivariate method is used to estimate the remaining background. Using these methods, the iron spectrum was measured in the energy range from 20 TeV to 500 TeV.}

\FullConference{35th International Cosmic Ray Conference --- ICRC2017\\
		10--20 July, 2017\\
		Bexco, Busan, Korea}

\begin{document}

\section{Introduction}
The energy spectrum of cosmic rays can be approximated by a power-law over many orders of magnitude, with several prominent features, most importantly the \emph{knee} (steepening) at around \SI{4}{\peta\eV} \cite{Kulikov_Khristiansen_1958,Hoerandel:2002yg}, the \emph{ankle} (flattening) above about \SI{4}{\exa\eV} \cite{Horandel:2006jd}, and a final cutoff at around \SI{40e18}{\eV} \cite{Settimo:2012zz}. Cosmic rays below the knee are commonly assumed to be accelerated within the Galaxy, possibly by supernova remnants (SNRs). Due to deflection by Galactic magnetic fields, the arrival directions of cosmic rays do not point back to their sources. Features in the energy spectrum can help solve the question of the origin of cosmic rays. 

The elemental spectra roughly follow the same overall power-law shape as the all-particle spectrum, with the location of the knee shifted to higher energies for heavier nuclei. Additional features (spectral hardening above a few hundred \si{GeV}) have been identified in the proton and helium spectra \cite{2011Sci...332...69A,PhysRevLett.114.171103,PhysRevLett.115.211101}. There are indications for similar features in heavier elements \citep{2010ApJ...714L..89A}. These features could be explained by nearby sources \cite{refId0}, multiple strong shocks in the accelerating SNRs \cite{0004-637X-763-1-47}, or propagation effects in the Galaxy \cite{2041-8205-752-1-L13}.  

Iron is the most abundant of the heavy elements in cosmic rays at \si{\TeV} energies. Characterising the iron spectrum is important for constraining the physics behind the spectral features, which in turn is important to characterize the origin of cosmic rays. The iron spectrum has been measured from tens of \si{GeV} almost to \si{\exa\eV} energies (cf. \cref{fig:ironspectrum}). The range from \SIrange{10}{1000}{\TeV} is not well covered by either direct detection methods or air shower arrays due to limited statistics for the former and poor charge resolution for the latter. 

Here, we present a measurement of the iron spectrum up to \SI{500}{\TeV} by the VERITAS experiment. For the first time, a template-based likelihood fit is adapted for the reconstruction of iron-induced showers. This improves the reconstruction of both the energy and the event geometry (arrival direction and core position). More details about the analysis presented here can be found in \cite{myThesis,myPaper}

\section{Instrument and Data Selection}
The VERITAS array\cite*{veritas,naheeperformance} comprises four IACTs, located at the Fred Lawrence Whipple Observatory (FLWO) in southern Arizona (31 40N, 110 57W,  1.3km a.s.l.). Each telescope has a camera covering a field-of-view of about \ang{3.5} diameter and consisting of 499 photomultiplier tubes (PMTs). VERITAS is designed to detect photons with energies between \SI{80}{\GeV} and more than \SI{30}{\TeV} in its current configuration.

VERITAS generally observes $\gamma$-ray sources or source candidates for between 15 minutes and several hours at a time. Even for strong sources, the rate of recorded events is dominated by cosmic-ray initiated showers. These events, which make up the main background for $\gamma$-ray studies with VERITAS, are utilized in this study to measure the cosmic-ray iron spectrum.

The VERITAS array has undergone two major upgrades: In 2009, the array layout was improved, improving the angular reconstruction\cite{2009arXiv0912.3841P}. PMTs with higher quantum efficiency were installed in all cameras in 2012, improving the array's response to low-energy showers in particular\cite{Kieda:2013wom}. In this paper, only data recorded between 2009 and 2012 were used. 71 hours of data taken in winter-like months during clear, moonless nights, with all four telescopes operating and an average elevation of \ang{80} or more were selected for the analysis. The strict selection criteria were necessary to ensure that simulations could be produced with sufficient statistics. The zenith angle cut in particular was chosen because particles from overhead have a higher chance of emitting DC light than particles with a larger zenith angle.

\section{Data Analysis}
The recorded showers are reconstructed using a template-based likelihood fit described below. The starting values for the fit are based on a geometric reconstruction, similar to the method described in \cite{veritasanalysis}. Dedicated energy lookup tables were produced using iron shower simulations. Only events with successful core and direction reconstruction and at least 70 hit pixels in each camera were selected for the likelihood fit, which reduces the data sample by a factor of about 50.

\subsection{Template Likelihood Reconstruction}
The standard geometric reconstruction is based on a handful of image parameters such
as the total signal in all pixels or the orientation of the image. This works well for $\gamma$-ray showers, which tend to have smooth, elliptical images. However, more advanced methods have been developed which take into account the signal in each pixel.  For example, a template-based likelihood fit was developed by the CAT experiment to reconstruct of gamma-ray induced showers \cite*{cattemplate} and is now used by other experiments as well \cite{Parsons201426,frogs}. A similar approach has been employed  in which the image templates have been replaced by a semi-analytic description of the air showers images \cite{deNaurois2009231}. 

For the template-based analysis, the recorded images are compared to the predictions from a library of template images, and the image parameters (energy and direction of the primary particle as well as the core position on the ground) that best describe the recorded image are selected. Details of the analysis can be found in \citep{myThesis}. The resolution obtained with this method depends mainly on the primary energy. The relative energy resolution is between \SIrange{6}{16}{\percent} after the selection cuts described in the next section. 

\subsection{Signal Selection and Direct Cherenkov Light}

The cosmic-ray events are dominated by showers induced by proton and Helium. To reduce this background, we exploited \emph{direct Cherenkov} (DC) light, which is emitted by charged primary particles before the first interaction \cite{Kieda}. As the nuclei emit coherently, the intensity of the DC light is proportional to the square of the nuclear charge. DC contribution to the camera image was identified by selecting pixels with high DC quality factor, defined as the ratio between the pixel's charge and the average charge in the neighboring pixels. For more details on the identification of DC contribution see \cite{myThesis,myPaper}. Only events where two or more out of the four cameras had a contribution from DC light were selected.

\subsection{Background Subtraction}
\begin{figure}[b]
\begin{center}
\includegraphics[width=0.6\textwidth]{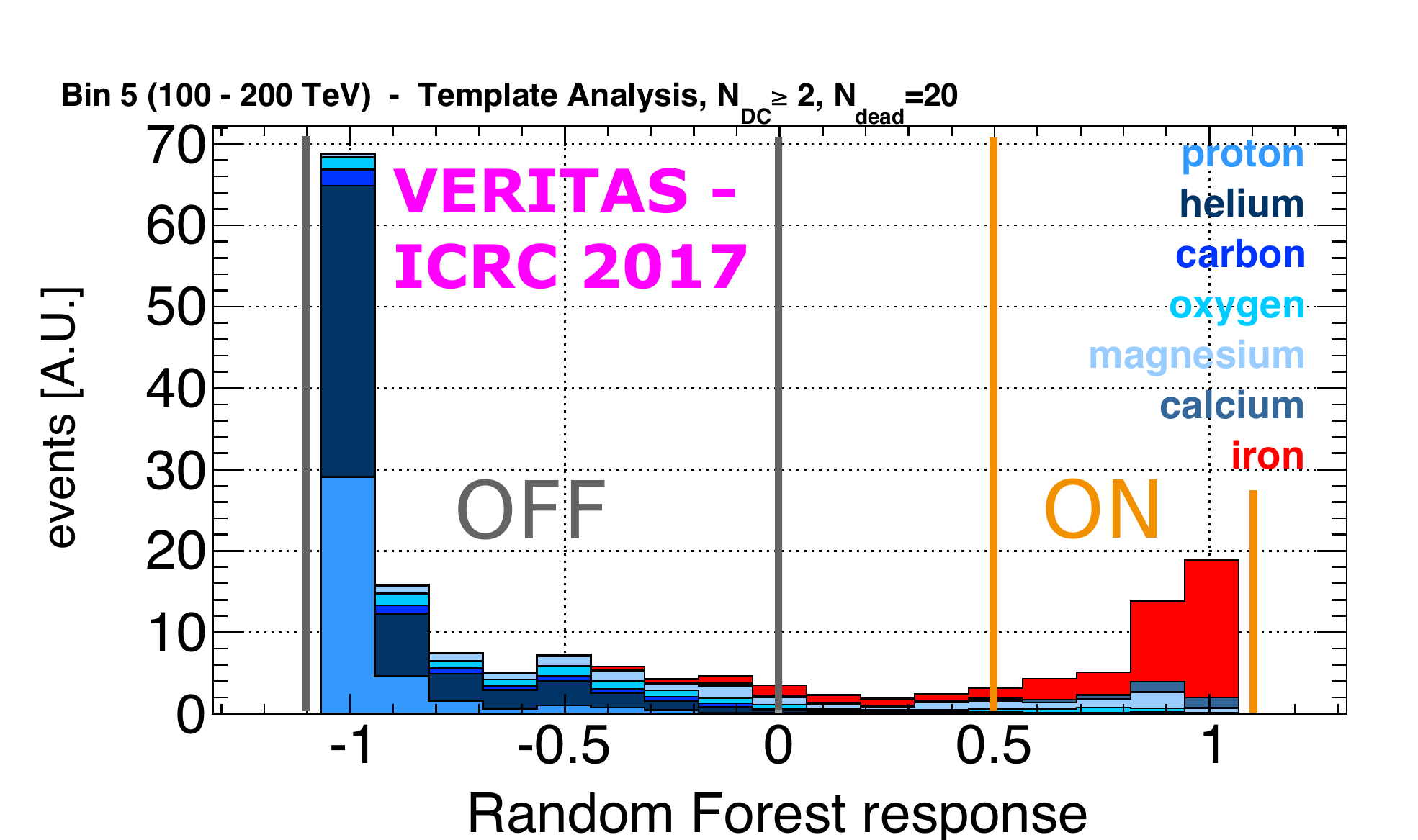}
\end{center}
\caption[Estimation of Signal Content Using Random Forest Response]{Random Forest classifier response (stacked histograms) in the energy bin from \SIrange{100}{200}{\TeV}.}
\label{fig:OnOff}
\end{figure}

Even after selecting events with DC pixels in at least two cameras (``DC sample''), a sizeable background due to lighter elements remains. In the case of protons and Helium nuclei, fluctuations and substructure in the shower can mimic DC light. Heavier elements such as magnesium or calcium emit a measurable amount of DC light themselves. 

To further suppress the background from lighter elements, Random Forest classifiers were trained on simulated cosmic-ray showers. An example for the distribution of the random forest response in the energy bin from \SIrange{100}{200}{\TeV} can be seen in \cref{fig:OnOff}. To estimate the number of iron and background events in the DC-sample, we selected an ON (signal-enriched) and OFF (background-enriched) region in the random forest response. The number of iron events $S$ in the DC sample is then estimated as 

\begin{align}
S = \frac{1}{\beta_s} \cdot \frac{N_{on} - \frac{\alpha_b}{\beta_b}\cdot N_{off}}{\frac{\alpha_s}{\beta_s} - \frac{\alpha_b}{\beta_b}}
\end{align}

Where $N_{on}$ and $N_{off}$ are the number of data events in the ON and OFF regions, respectively, and $\alpha$ and $\beta$ are the relative selection efficiencies for ON events and OFF events, obtained from simulations: 

\begin{align}
\alpha_s &= \frac{S_{on}^{MC}}{S^{MC}} 
& \alpha_b &= \frac{B_{on}^{MC}}{B^{MC}} \nonumber \\
\beta_s &= \frac{S_{off}^{MC}}{S^{MC}} 
& \beta_b &=   \frac{B_{off}^{MC}}{B^{MC}} \label{eq:alphabeta}
\end{align}

Assuming Poissonian uncertainties on the counts, the statistical uncertainty on the number of signal events is given by

\begin{align}
\Delta S &= \sqrt{\left( \frac{\partial S}{\partial N_{on}} \cdot \Delta N_{on} \right)^2 + \left( \frac{\partial S}{\partial N_{on}^{data}} \cdot \Delta N_{on}^{data} \right)^2 } \nonumber \\
	&=  \left( \frac{\beta_b}{\alpha_s\cdot \beta_b - \alpha_b\cdot \beta_s} \right) \cdot \sqrt{N_{on} + \frac{\alpha_b^2}{\beta_b^2}N_{off}}.
\end{align}

\section{Results}
\begin{table*}[tb]
\begin{tabular*}{\textwidth}{@{\extracolsep{\fill} } lrrrrrrrr}
\toprule

Bin & $E_{min}$ & $E_{max}$ & $E_c$ & $N$ & $N_{on}$ & $N_{off}$ & $S$~~ & differential flux \\ 
  & [\si{\TeV}] & [\si{\TeV}] & [\si{\TeV}] & & &  & & [\si{\per\metre\squared\per\second\per\TeV\per\steradian}] \\ 
\midrule 
0 & 20 & 25.1	& 22.4 & 192 & 75 &  102 & \num[separate-uncertainty=true]{127 \pm 19}\phantom{.} & \num[scientific-notation=true, separate-uncertainty=true]{5.8 \pm  0.9 e-06} \\
1 & 25.1 & 31.6	& 28.2 & 189 & 76 & 105 & \num[separate-uncertainty=true]{86\pm 15}\phantom{.} & \num[scientific-notation=true, separate-uncertainty=true]{2.1 \pm 0.4 e-06} \\
2 & 31.6 & 39.8	& 35.5 & 171 & 51 & 103 & \num[separate-uncertainty=true]{65 \pm 13}\phantom{.} & \num[scientific-notation=true, separate-uncertainty=true, zero-decimal-to-integer=false]{1.0 \pm 0.2 e-06} \\
3 & 39.8 & 50.1	& 44.7 & 147 & 45 & 95 & \num[separate-uncertainty=true]{55 \pm 11}\phantom{.} & \num[scientific-notation=true, separate-uncertainty=true]{6.1 \pm 1.2 e-07} \\
4 & 50.1 & 100.	& 70.8 & 337 & 67 & 229 & \num[separate-uncertainty=true]{71 \pm 12}\phantom{.} & \num[scientific-notation=true, separate-uncertainty=true]{1.5 \pm 0.3 e-07} \\
5 & 100. & 200.	& 141.3 & 197 & 41 & 141 & \num[separate-uncertainty=true]{37 \pm 8}\phantom{.0} & \num[scientific-notation=true, separate-uncertainty=true, zero-decimal-to-integer=false]{4.0 \pm 0.8 e-08} \\
6 & 200. & 500. & 316.2 & 65 & 13 & 48 & \num[separate-uncertainty=true]{8.8 \pm 4.4} & \num[scientific-notation=true, separate-uncertainty=true]{3.1 \pm 1.6 e-09} \\
\bottomrule
\end{tabular*} 
\caption[Differential Flux of Cosmic Ray Iron Nuclei]{Events and differential flux in each energy bin. $E_c$: logarithmic bin center. $N$: total number of events after analysis cuts. $N_{on}$, $N_{off}$: counts in the ON and OFF regions (cf.  \cref{fig:OnOff}), $S$: derived number of signal (iron) events.}
\label{tab:ironspectrum}
\end{table*}

\begin{figure*}[b]
\subfloat[][Differential flux.]
{\includegraphics[width=0.5\textwidth]{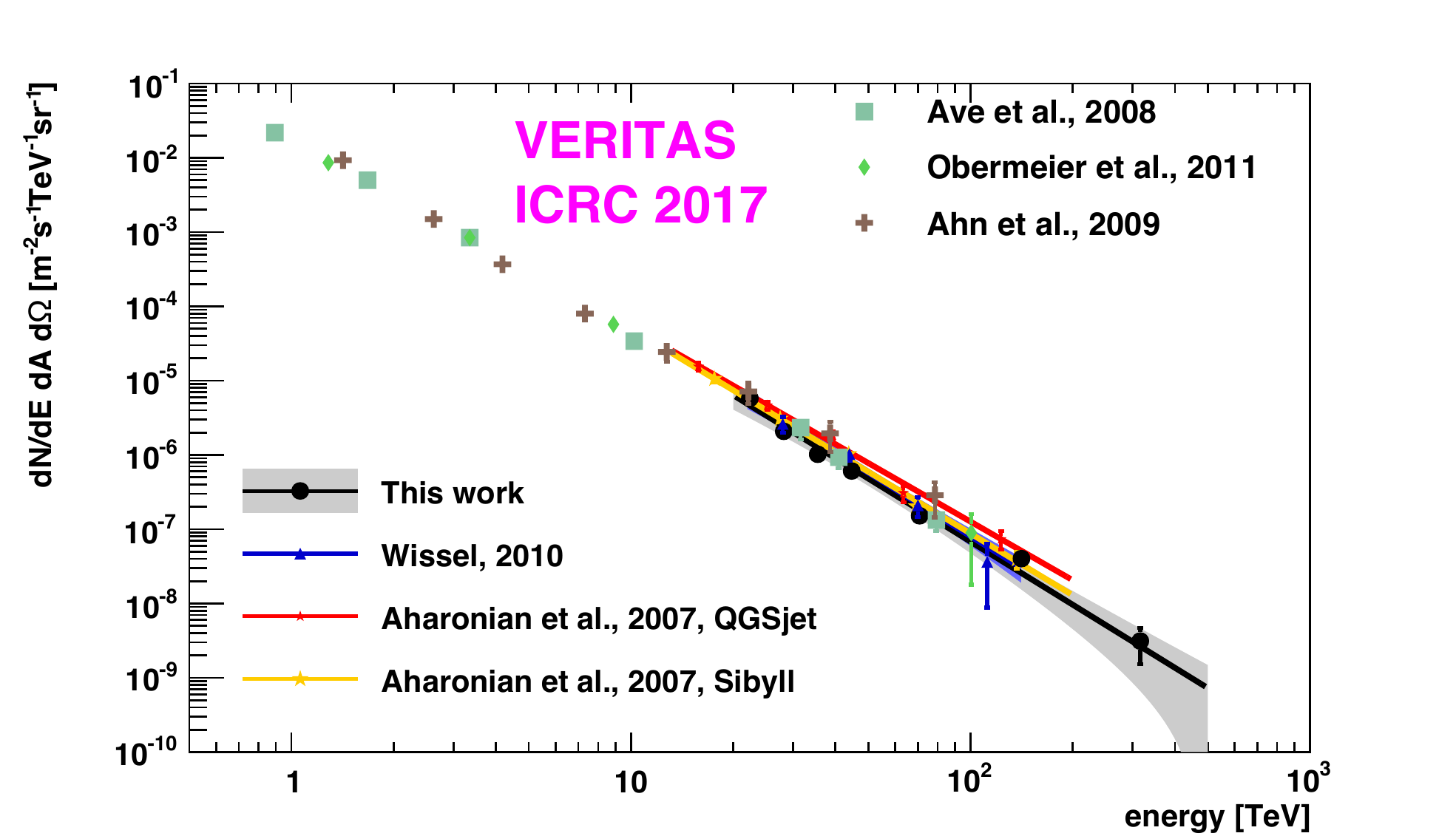}}
\subfloat[][Differential flux multiplied by $E^{2.7}$.]
{\includegraphics[width=0.5\textwidth]{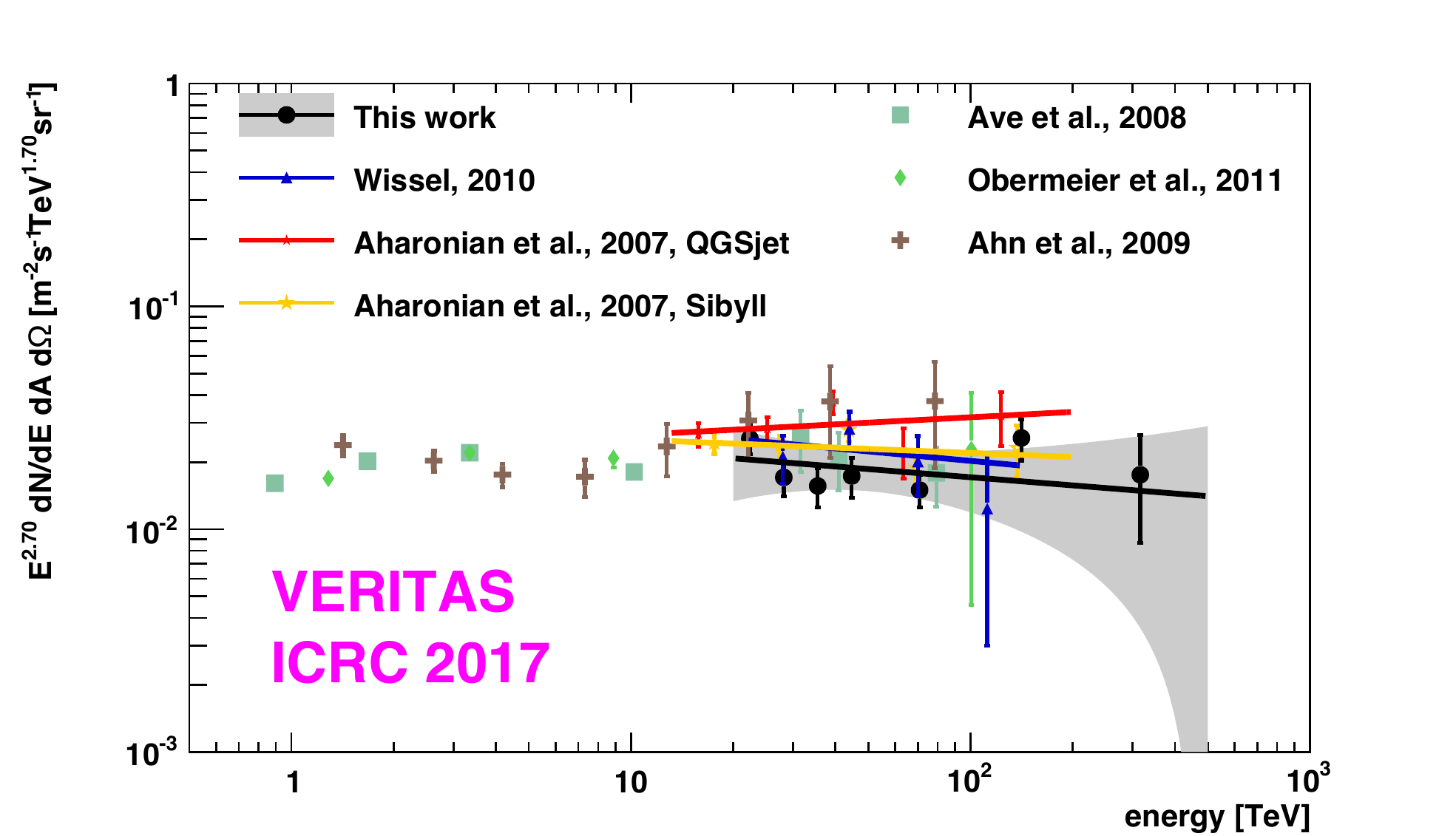}}
\caption[The Cosmic-Ray Iron Spectrum]{Iron spectrum in cosmic rays from the study presented here. Previous measurements by VERITAS \citep{wissel}, H.E.S.S. \citep{hess}, TRACER \cite*{Ave:2008as,2011ApJ...742...14O} and CREAM \citep{0004-637X-707-1-593}. Statistical uncertainties only.}
\label{fig:ironspectrum}
\end{figure*}

The 71 hours of VERITAS data passing all selection criteria were reconstructed using the template likelihood method. The results were binned in energy, with the bin sizes chosen to have a roughly flat distribution of event per bin, except for the last bin. Finally, the random forest classifiers were applied to the resulting events. \Cref{tab:ironspectrum} shows the total number of events and the number of iron events per bin, as well as the differential flux. The spectrum (plotted in \Cref{fig:ironspectrum} is well-fit by a power law

\begin{align}
\frac{\de N}{\de E \de A \de t \de \Omega} = f_0\cdot \left(\frac{E}{E_0}\right)^{-\gamma}
\label{eq:PL}
\end{align}
over the whole energy range with normalization energy $E_0=\SI{50}{\TeV}$, normalization factor $$f_0 = ( 4.82 \pm 0.98_{stat}\,^{+2.12}_{-2.65 sys} )  \cdot \num[scientific-notation=true, exponent-to-prefix=false,retain-unity-mantissa = false]{1e-7} \si{\per\metre\squared\per\second\per\TeV\per\steradian} $$ and index $$\gamma = 2.82 \pm 0.30_{stat}\,^{+0.23}_{-0.25 sys}.$$

The list of systematic uncertainties on the flux normalization and index can be found in \cref{tblSystComb}. 

The results agree well with previous measurements by IACTs and direct detection experiments \citep{wissel,hess,Ave:2008as,2011ApJ...742...14O,0004-637X-707-1-593}. The spectrum was extended beyond \SI{200}{\TeV}, with no indication of a cutoff. The template fit was able to compensate for the truncation of high-energy events, whose images are not fully contained in the IACT cameras, which increased the maximum energy of the analysis compared to previous measurements by IACTs \cite{wissel,hess}.

\begin{table*}[tb]
\begin{tabular*}{\textwidth}{@{\extracolsep{\fill} } lrr}
\toprule
 Cause &  Effect on $f_0$  &  Effect on $\gamma$  \\
  \midrule
 Absolute calibration (incl. atmosphere, detector model)  & $\pm \SI{40}{\percent}$ & $\pm 0.2$ \\
 `Dead' pixels (broken/turned off due to starlight)  & $\pm \SI{7}{\percent}$ & $\pm 0.07$ \\
 Intrinsic energy bias  & $^{+\SI{0}{\percent}}_{-\SI{30}{\percent}}$ & $^{+0.0}_{-0.1}$ \\ 
 Statistical uncertainty on effective area & $\pm \SI{10}{\percent}$ & --- \\ 
 Hadronic interaction model & $\pm \SI{12}{\percent}$ & $\pm 0.1$  \\
 Remaining background & $^{+\SI{0}{\percent}}_{-\SI{15}{\percent}}$ & ---  \\
 \midrule
  Total & $^{+\SI{44}{\percent}}_{-\SI{55}{\percent}}$ &  $^{+0.23}_{-0.25}$\\
 \bottomrule
\end{tabular*}  
\caption[Systematic Uncertainties of the Cosmic-Ray Iron Spectral Parameters]{Systematic uncertainties of the cosmic-ray iron spectral parameters}
\label{tblSystComb}
\end{table*}
\section{Future Outlook}
There are several concievable ways in which the results presented here could be extended and improved upon in future studies. Including the first interaction height in the likelihood fit is expected to further improve the energy resolution, and would also provide a good cross-check of hadronic and nuclear interaction models. Compared to VERITAS, some of the telescopes in the upcoming CTA array \cite{2013APh....43....3A}  will have larger mirror area and finer pixels. CTA should be able to measure the iron spectrum with better resolution and down to lower energies, and may be sensitive to possible features in the iron spectrum, such as spectral hardening.

Additionally, it is possible to extend the template library to lighter elements. Using a combination of image parameters, DC light, and goodness-of-fit, it should be possible to measure the composition of cosmic rays in the \si{\TeV} to \si{\PeV} range.

\section{Summary and Conclusions}
In this study, we have successfully applied the template-based likelihood reconstruction to measure the cosmic-ray iron spectrum in the energy range from \SI{20}{\TeV} to \SI{500}{\TeV}. The resulting spectrum follows a power-law shape and agrees well with previous measurements by IACTs and direct detection experiments. Due to the energy resolution and the systematic uncertainties on the energy scale, we cannot exclude possible spectral features similar to the hardening of $\Delta\gamma\approx 0.1$ observed by AMS in the proton and helium spectra.

In the future, next-generation IACT arrays such as the planned CTA observatory will be able to expand upon the measurements presented here, e.g. extending the energy range or measuring other elemental spectra.

\section*{Acknowledgments}
VERITAS is supported by grants from the U.S. Department of Energy Office of Science, the U.S. National Science Foundation and the Smithsonian Institution, and by NSERC in Canada. We acknowledge the excellent work of the technical support staff at the Fred Lawrence Whipple Observatory and at the collaborating institutions in the construction and operation of the instrument. The VERITAS Collaboration is grateful to Trevor Weekes for his seminal contributions and leadership in the field of VHE gamma-ray astrophysics, which made this study possible. H. Fleischhack gratefully acknowledges support through the Helmholtz Alliance for Astroparticle Physics.

\setlength{\bibsep}{0pt plus 0.3ex}
\bibliographystyle{unsrt}
\bibliography{bibliography.bib}


\end{document}